\documentstyle[aps,prl,psfig,epsfig,preprint,tighten]{revtex}

\begin{document}

\title{Use of optically detected magnetic resonance to correlate germanium
electron centers with UV absorption bands in X-ray irradiated germanosilicate glasses}

\draft

\author{D. Poulios, N. P. Bigelow, and J. P. Spoonhower}
\address{Department of Physics and Astronomy, University of Rochester,}
\address{Rochester, New York 14627}

\date{\today}

\maketitle

\begin{abstract}

The relationship between paramagnetic defect centers and UV
absorption bands simultaneously generated by ionizing radiation in
Ge-doped SiO$_{2}$ glass is investigated using magneto-optical
techniques.  A sample of 7.0 mol$\%$ Ge-doped SiO$_{2}$ was
exposed to X-ray radiation, which resulted in the formation of
absorption bands centered at 4.4 eV and 5.7 eV as well as electron
spin resonance (ESR) signals attributed to the Ge(1) and Ge(2)
defects. To isolate paramagnetic contributions to the induced
optical absorption, the magnetic circular dichroism absorption
(MCDA) spectrum was measured at several different temperatures
over the range 2.00-6.21 eV.  The optically detected magnetic
resonance (ODMR) spectrum was then obtained at 4.42 eV via
microwave-induced changes in the MCDA signals in order to
unambiguously correlate this optical transition with a particular
ESR signal.  The ODMR data indicates that the Ge(1) center is
responsible for this absorption band, providing the first
unequivocal correlation of ESR and optical properties for this
defect.  In addition, evidence of a weakly absorbing paramagnetic
defect was found at 5.65 eV that appears to be related to the
Ge(2) center.

\end{abstract}



\section{Introduction}

Germanosilicate glasses and germanium-doped silica fibers have
been extensively studied for over a decade, mostly because of
their photosensitive properties and utility in producing Bragg
gratings and other optical devices.  Despite the intense research
efforts, however, the precise mechanisms responsible for the
photosensitive effect remain unclear.  Many researchers have
focused on elucidating the defect photochemistry in the glass
structure as a possible means of explaining this effect.  In the
most widely accepted model, photorefractive index changes in
Ge-doped SiO$_{2}$ induced by exposure to ultraviolet light or
ionizing radiation are brought about by the presence of germanium
oxygen-deficient defect centers (GODCs) in the glass network
\cite{Atkins}. Two such defects thought to play significant roles
in the photosensitive phenomenon are the neutral oxygen
monovacancy (NOMV) and the neutral oxygen divacancy (NODV or
Ge$^{+2}$), both of which have electronic transitions in the
vicinity of 5.16 eV \cite{Hosono}. NOMV and NODV are known to be
photobleached by UV radiation via one and two-photon effects
\cite{Nishii,Fujimaki}, respectively, although the exact mechanism
by which the NODV bleaching occurs is subject to considerable
debate \cite{Nishii,Fujimaki,Essid,Fujimakitwo}. These
photobleaching processes are known to lead to the production of
several broad UV absorption bands, the most notable of which are
centered at 6.5 eV, 5.8 eV, and 4.5 eV
\cite{Essid,Neustruev,Shigemura}.

Also produced in these processes are a number of paramagnetic
point defects which may play a significant role in UV-induced
index changes, the most relevant being the two varieties of
germanium trapped electron centers (GECs), known as Ge(1) and
Ge(2), and the GeE$'$ center.  Ge(1) is believed to be a
substitutional Ge tetrahedron with four silicon atoms as the next
nearest neighbors \cite{Tsai} that has trapped an electron donated
by either an NODV or a bridging oxygen \cite{Nishii,Fujimaki}.
Various experiments have shown that this defect center can be
produced by either two-photon absorption of excimer laser light
(KrF, XeCl, or ArF) \cite{Nishii,Fujimaki,Nishiitwo}, single
photon absorption of light from a KrCl excimer lamp
\cite{Fujimakitwo}, or by ionizing radiation such as X- or
$\gamma$-rays \cite{Tsai,Friebele}. A similar structure has been
proposed for the Ge(2) defect (electron trapped at Ge tetrahedron
with one Ge and three Si atoms as next nearest neighbors)
\cite{Tsai}, although there is significant recent evidence to
suggest the Ge(2) is actually an NODV which has trapped a hole
\cite{Fujimakitwo,Anoikin}.  The GeE$'$ center, consisting of a
threefold coordinated Ge with an unpaired electron, is readily
produced by single photon scission of the metal-metal bond of an
NOMV \cite{Hosono,Hosonotwo}; there is also evidence that GeE$'$
can be formed by the photobleaching of GECs \cite{Nishii}.

Assignment of these defects to particular optical absorption
features has proven difficult for several reasons.  One is the
somewhat formless nature of the optical absorption spectrum of
irradiated germanosilicate glasses caused by the overlap of the
induced absorption bands with the GODCs.  This overlap makes
accurate spectral separation of absorptions quite difficult.
Electron spin resonance (ESR) spectroscopy, often of great use in
identifying defects in solids, suffers a similar difficulty when
the GECs and the GeE$'$ center are simultaneously present, as
their spectra are considerably entangled.  In an effort to
correlate these germanium-related paramagnetic species with
induced absorption bands, concomitant measurements of ESR and
optical absorption spectra have been conducted as a function of
annealing temperature.  This technique has been used to attribute
the 6.5 eV, 5.8 eV, and 4.5 eV bands to GeE$'$, Ge(2), and Ge(1)
\cite{Friebeletwo,Neustruevtwo}, respectively, but it must be
noted that it does not provide unambiguous correlation between a
given defect's ESR and absorption spectra, particularly when
multiple centers with broad overlapping bands are present.

To correlate these properties in a direct manner for the Ge(1) and
Ge(2) centers, we first measured the magnetic circular dichroism
absorption (MCDA) spectrum of an X-ray irradiated 7.0 mol$\%$
Ge-doped SiO$_{2}$ sample over the range 2.07-6.21 eV at various
temperatures to establish which bands are paramagnetic.  This
technique also provided spectral separation of the induced
paramagnetic bands from those of the GODCs and other diamagnetic
defects.  We then examined the optically detected magnetic
resonance (ODMR) spectrum associated with MCDA signal extrema via
microwave-induced changes in the MCDA signal.  Using this ODMR
technique, we were able to establish unequivocal correlations
between optical transitions and specific ESR spectra; this has not
been accomplished previously. Identification of the defects seen
in the ODMR spectra was possible by making comparisons with ESR
data taken on a conventional spectrometer.

\section{Experiment}
\subsection{Sample Preparation}
The germanosilicate glass used in the experiment was hewn from a
section of 7.0 mol$\%$ GeO$_{2}$, 93.0 mol$\%$ SiO$_{2}$
single-mode fiber preform fabricated using the vapor axial
deposition (VAD) method. The preform was first cut into a 10 mm x
5 mm x 2 mm slab and polished on two sides. According to the glass
specifications, the germanium concentration varied by less than
1$\%$ from the center to the edges of the sample.  The sample was
then exposed to radiation from a 2.4 kW tungsten anode X-ray tube
(50 kV, 48 mA) at room temperature in order to create defects
throughout the glass network.  The defect concentration was
periodically measured through double numerical integration of ESR
spectra and comparison to a standard sample containing a known
number of Si E$'$ centers.  Irradiation was halted once the
induced spin density of the germanosilicate sample became
saturated (~10$^{17}$ spins/cm$^{-3}$).

\subsection{MCDA and ODMR Theory}

MCD is an effect common to all matter whereby the application of a
static uniform magnetic field to a material induces an asymmetry
in the optical densities for right- and left-hand circularly
polarized light (hereby referred to as A$_{+}$ and A$_{-}$,
respectively) propagating parallel or antiparallel to the field.
In simplest terms, the MCD phenomenon can be viewed as a sum of
three separate manifestations of the Zeeman Effect:  (1)
unresolved field-induced splitting of a degenerate ground or
excited state, (2) mixing of nearby states by the magnetic field,
and (3) population differentials between ground state magnetic
sublevels \cite{Geschwind}.  The first two contributions, often
grouped into the single term "diamagnetic effect", are
temperature-independent and linear in field strength; the third
term depends strongly upon the temperature and is referred to as
the paramagnetic term.  It is then possible to write the MCDA
signal as
\begin{equation}
\Delta A = \Delta A _{para} (B,T) + \Delta A _{dia} (B),
\end{equation}
where $\Delta$A$_{para}$ is the paramagnetic term,
$\Delta$A$_{dia}$ is the diamagnetic term, B is the magnetic field
strength, and T is the temperature \cite{Spaeth}.  For systems
with degenerate ground states, the paramagnetic term tends to
dominate the MCDA spectrum, particularly at low temperatures.  Use
of low temperatures in MCDA spectroscopy is especially
advantageous when a system has a complicated optical absorption
spectrum consisting of many overlapping bands, and one wishes to
differentiate paramagnetic contributions from diamagnetic ones.

The paramagnetic part of the MCDA signal is also useful for the
detection of magnetic resonance in the ground state.  For
S$_{1/2}$ systems (such as the GECs), the differential absorption
may be written as the simple expression
\begin{equation}
\Delta A \propto Tanh \left( \frac{g \mu _{B} H}{2 k T} \right)
\end{equation}
where it is assumed that the equilibrium population distribution
amongst the ground state Kramers doublet follows Boltzmann
statistics, and that the diamagnetic contributions to the MCDA
signal may be ignored \cite{Stephens}.  Equation (2) indicates
that MCDA can be used as a sensitive probe of the spin temperature
of a given system's ground state.  Observation of ODMR is then
carried out by measuring perturbations in the MCDA signal (and,
hence, the spin temperature) brought about by driving spin
transitions between ground state sublevels with resonant
microwaves of sufficient power to overcome the spin-lattice
relaxation rate. This optical method of detecting spin resonance
transitions offers several advantages over conventional ESR
techniques, the most significant being that optical rather than
microwave quanta are used in the detection process (providing up
to a $\sim$10$^{4}$ gain in sensitivity), and that the optical
properties of a given defect are unambiguously correlated with its
spin resonance spectrum.  It should be noted that the selectivity
of the ODMR method allows for the disentanglement of the optical
and ESR properties of simultaneously present defects provided only
that these properties are not identical.

\subsection{Experimental Apparatus and Techniques}

MCDA spectra were recorded at temperatures of 4.2 K and 1.6 K. The
MCD spectrometer used consisted of a 30 W deuterium lamp focused
onto the entrance slit of a single grating 0.25 m monochromator
having a bandwidth of less than 3 nm over the range 2.07-6.21 eV.
The output light was linearly polarized by a Glan-Taylor
polarizing prism and directed through a 42 kHz photoelastic
modulator, which alternately produced right- and left-hand
circularly polarized light at its fundamental frequency. The
sample was mounted in a variable-temperature helium cryostat with
four-way optical access outfitted with a superconducting magnet
capable of operating at 1.8 T with a homogeneity of 1 part in
10$^{4}$ over 1 cm$^{3}$.  The polarization-modulated light was
then focused onto the sample by a set of computer controlled
motorized lenses designed to account for chromatic aberration.
Light transmitted by the sample was then trained on a
photomultiplier tube connected to a feedback circuit which varied
the tube voltage in order to keep the DC light signal constant as
a function of wavelength.  A lock-in amplifier referenced to the
modulator frequency measured the differential absorption of the
two polarizations of light; the ratio of the lock-in signal to the
DC light signal yields a quantity proportional to  , the
difference in optical densities for right and left circularly
polarized light \cite{Drake}.  A computer scanned the spectrometer
in 0.5 nm steps and recorded the lock-in voltage, DC light signal,
and tube voltage in order to obtain the full MCDA spectrum.  A
long pass filter was used for the range 3.45-6.21 eV to filter out
second order UV light from the monochromator.

Optical absorption spectra in the range 2.26-6.21 eV were measured
at room temperature before and after irradiation using a
commercial 2-beam spectrophotometer.  The spectral bandwidth was
kept at 3 nm, and the spectra were adjusted for instrument
response by baseline subtraction.

To obtain ODMR spectra below 5.18 eV, the deuterium
lamp-monochromator system was replaced by a 150 W high-pressure
mercury lamp equipped with a 0.1 m monochromator for improved
light throughput.  Unmodulated 24.1 GHz (K-band) microwaves from a
600 mW klystron were then coupled into a cylindrical microwave
sample cavity using a coaxial antenna. As the magnetic field was
scanned, the computer recorded negative changes in the MCDA signal
caused by resonant microwave-driven  transitions between ground
state Zeeman sublevels as a function of field strength.
Acquisition of ODMR spectra above 5.18 eV was carried out in much
the same manner, the only difference being that a deuterium lamp
filtered by a 220 nm interference filter was used as a light
source to compensate for the monochromator's weak light throughput
in the deep UV.  The bandwidth of the interference filter was 10
nm and transmitted approximately 0.1$\%$ of incident light at 5.28
eV and 6.27 eV.  All ODMR spectra were acquired at a sample
temperature of 1.6 K.  For purpose of comparison, ESR spectra were
also recorded on a separate commercial unit at 300 K using 9.6 GHz
(X-band) microwaves at power level of 13.6 mW and a field
modulation of 100 kHz.  The signal amplitude and spin density were
measured at several different microwave power levels to ensure
that the sample was not being saturated.  The magnetic field
sweeps for both the ESR and ODMR spectrometers were calibrated
using diphenylpicrylhydrazyl (DPPH, g=2.0036) as a standard.

\section{Results and Discussion}
\subsection{ESR Results}

Prior to X-ray exposure, the Ge-doped glass sample shows no
measurable ESR activity.  After 15 hours of X-irradiation at 300
K, an ESR signal (shown in figure 1) emerges which closely
resembles the spectra of the GECs previously reported in the
literature \cite{Fujimakitwo,Friebele,Fujimakithree,Chiodini}. The
spectrum shown in figure 1 contains overlapping ESR contributions
from both the Ge(1) and Ge(2) centers.  In an effort to resolve
the composite irradiated germanosilicate glass ESR spectrum into
its individual components, researchers have conducted various
isochronal annealing experiments.  Several results emerge from
these studies: One is the correlation of particular features of
the GEC ESR signal with individual defects, such as the local
minimum at g=1.9934, attributed to the Ge(1) defect, and the
minimum at g=1.9869, related to the Ge(2) center
\cite{Friebele,Fujimakithree}.  Another significant finding is
that there is some overlap between the Ge(1) and Ge(2) signals
above g=2.000 \cite{Guryanov} but the two are otherwise spectrally
separate. It is interesting to note that the typical ESR pattern
of the GeE$'$ center \cite{Hosono,Friebele} is absent from figure
1, indicating that the X-irradiation did not photochemically
produce this defect in any measurable quantity. This finding
contrasts with experiments conducted with excimer laser light,
where it has been suggested that GECs are produced in the initial
stages of irradiation and subsequently transformed to GeE$'$
centers with continued exposure \cite{Nishii}.

\subsection{Optical Absorption and MCDA}

The optical absorption spectra of the glass sample prior to
X-irradiation is shown in figure 2a, with the only notable detail
being the strong GODC absorption band centered at 5.16 eV.  Also
shown in figure 2a is the post-irradiation absorption spectrum. As
is evident in the plot, it is difficult to ascertain the precise
shape and location of the peaks of the radiation-induced bands due
to their considerable overlap with the broad NODV and NOMV optical
transitions.  The strategy most frequently employed to obtain
these quantities in irradiated Ge-doped SiO$_{2}$ systems is to
fit multiple gaussian peaks to the induced absorption spectrum, as
is shown in figure 2b.  Since no evidence of the GeE$'$ center was
seen in the ESR data, the induced absorption spectrum was fit with
three gaussian functions (representing Ge(1), Ge(2), and the
GODCs) rather than the usual four.  In this figure, the best fit
to the data shows the growth of absorption bands centered around
4.73 eV and 6.06 eV, while the initial 5.16 eV absorption band is
depleted; these results are consistent with what is generally
found in the literature
\cite{Nishii,Fujimaki,Essid,Fujimakitwo,Shigemura,Essidtwo}. While
this fitting procedure is useful in that it provides a rough
estimate of the optical properties of the induced defects, the
process is obviously prone to considerable error, particularly the
choice of initial conditions for the fits. One must also account
for the possibility that one or more diamagnetic species could be
induced in the glass that would obviously be undetectable by ESR
spectroscopy; these defects would of course be ignored in the
fitting process.

MCDA spectroscopy is uniquely suited to circumvent this difficulty
as the Ge(1) and Ge(2) centers provide significant paramagnetic
contributions while the NOMV and NODV show only comparably weak
diamagnetic signals; thus, the transitions are easily spectrally
separated using their respective temperature dependencies.  The
MCDA spectra of the irradiated and unirradiated (inset) sample
measured at 1.6 and 4.2 K with a field strength of 1.8 T are
presented in figure 3. The spectra were corrected for diamagnetic
signal contributions from the fused  quartz cryostat windows as
well as for zero-field baseline effects.  Prior to irradiation,
the MCDA spectrum at 4.2 K is featureless except for a
negative-going band peaking at 5.16 eV that is clearly
attributable to singlet-singlet GODC transitions.  Lowering the
temperature to 1.6 K produces no change in the GODC MCDA signal
strength, providing further evidence to support the long-standing
assumption that the GODCs are diamagnetic in nature.  Despite a
careful search, no evidence of the singlet-triplet NODV transition
believed to occur at 3.8 eV \cite{Neustruev} could be found. This
could due to the band's weak oscillator strength as well as its
diamagnetic character.  Upon completion of the X-irradiation,
growth of an MCDA band peaking at 4.42 eV is observed, as well as
a slight decrease in the depth of the GODC signal, indicating that
these centers undergo a small degree of photobleaching.  It is
quite clear upon viewing the spectra that paramagnetic centers are
responsible for the 4.42 eV band, as the peak signal strength
increases by a factor of approximately 1.8 upon lowering the
temperature from 4.2 K to 1.6 K.  No other paramagnetic signals
were as readily apparent in figure 3, possibly due to the
spectrometer's low light throughput in the deep UV, which required
the use of greater photomultiplier tube voltages to produce
measurable signals.  This effect significantly degraded the
signal-to-noise ratio of the spectra above 5.50 eV and made it
difficult to discern any structure, and may also have obscured
weak absorption bands overlapping with the GODCs in this range.

\subsection{ODMR and Tagged MCDA Results}

The upper trace of figure 4 shows a plot of the ODMR spectra of
the X-irradiated sample measured at 4.42 eV.  The technique used
for obtaining ODMR spectra involved coupling fixed-frequency CW
microwaves into the sample cavity and monitoring the MCDA signal
as the field strength was increased; the signal dips in figure 4
indicate where a particular defect's ground state splitting is
resonant with the frequency of the microwave field.  The ODMR data
shows that a spin resonance signal peaking at g=2.000 is directly
associated with the induced absorption band at 4.42 eV, providing
conclusive evidence that the GECs are indeed responsible for at
least a portion of the induced absorption spectrum at this energy.
ODMR was also attempted at 5.65 eV to account for the possibility
that a defect with a weak paramagnetic absorption at this energy
might be obscured by the high signal noise and thus would not be
apparent in the MCDA spectra of figure 3.  As was mentioned
previously, a 220 nm (5.65 eV) interference filter was employed
rather than the monochromator to increase light throughput.  The
5.65 eV ODMR signal is shown in the bottom trace of figure 4.  It
is similar to the 4.42 eV trace in that it has a peak at g=2.000,
but also contains additional significant signal contributions
below g=1.990; these results indicate that the GECs also absorb in
the deep UV range transmitted by the filter.  There are several
possible explanations for the presence of the g=2.000 signal at
5.65 eV.  One is that the 4.42 eV band is broad enough to overlap
with the filter transmission range so that it "leaks" into the
5.65 eV ODMR signal, while another possibility is that the defect
responsible for the 4.42 eV band has another separate higher
energy transition.  Although several reports, including recent
studies of defect-free irradiated germanosilicate glasses by
Chiodini et al \cite{Chiodini}, have supported the latter
possibility, the current data does not permit exclusion of either
scenario.  It seems the most likely explanation for the additional
broadness of the 5.65 eV ODMR signal is that a defect unrelated to
the 4.42 eV band weakly absorbs in this region.  It is impossible
to deduce where the peak oscillator strength of this defect lies
based on the data presented here, but considering the relative
weakness of the MCDA signal in the range 5.0-6.2 eV, it is
reasonable to ascribe this defect to one of the broad absorption
bands located at higher energies(6.2-7.5 eV) found in deep UV
spectroscopy studies \cite{Neustruev}.  Part of the ODMR signal at
5.65 eV could then be directly related to the "tail" of one of
these bands found at higher energies.

The question of which defect is responsible for these optical
transitions can be addressed by comparing the ODMR data to the ESR
data of figure 1.  Because ODMR directly records the absorption of
microwave power as a function of magnetic field strength rather
than the corresponding derivative (as in conventional ESR
spectroscopy), direct comparisons to the ESR data are tedious
unless the latter is first integrated.  Figure 5a shows a
comparison of the integrated GEC ESR spectrum displayed in figure
1 to the ODMR spectrum measured at 4.42 eV shown in figure 4.  The
two techniques show excellent agreement, as it is clear from
figure 5a that both the ODMR and ESR spectra share a peak at
g=2.000 and have similar bandshapes despite the great difference
in temperature.  Perhaps the most significant aspect of this
comparison is the lack of a shoulder in the vicinity of g=1.985 in
the ODMR spectrum, suggesting that the Ge(2) center plays no role
in the optical transition at 4.42 eV; thus, the Ge(1) defect is
solely responsible for the radiation-induced paramagnetic
absorption here.  The ODMR results also indicate that the
composite ESR spectrum of the GECs above g=1.990 is dominated by
Ge(1).  This assertion is supported by the work of Anoikin et al,
who were able to isolate the Ge(1) ESR signal in Ce$^{+3}$-doped
germanosilicate glass; though they did not supply particular g
values for the Ge(1) center, the ESR spectrum shown in their work
closely resembles the composite GEC spectrum shown in figure 1
\cite{Anoikin}.  A direct comparison of the 5.65 eV ODMR spectrum
and the integrated GEC ESR spectrum is shown in figure 5b.
Although the poor signal-to-noise ration precludes rigorous
assessment of the features of the ODMR spectrum, intriguing
possibilities emerge from the comparison.  The signal dip at
approximately g=2.000 seems to indicate the presence of the Ge(1)
defect; however, the nonzero ODMR signal below g=1.990 invites
correlation with the Ge(2) shoulder of the integrated ESR
spectrum, a feature undetected in the 4.42 eV ODMR. Unfortunately,
lack of a clean ODMR signal prevents definite assignments here,
and further studies are clearly necessary to ascertain which
defects are present.

The ODMR results may also be used to further elucidate the optical
properties of the induced defects.  By tuning the magnetic field
to a particular spin resonance transition and measuring the MCDA
signal as the wavelength is varied, it is possible to trace out an
ODMR excitation spectrum for a given defect.  This method, known
as "tagged MCDA" \cite{Spaeth}, serves a number of purposes. It
identifies all optical transitions associated with a paramagnetic
defect and effectively isolates its individual MCDA spectrum from
the total MCDA, which is particularly useful when separating
overlapping bands belonging to different defects \cite{Spaeth}. It
is also possible to use tagged MCDA to estimate the peak position
and width of an absorption band, though it is not necessarily a
reliable measure of bandshape \cite{Spaethtwo}.  The tagged MCDA
spectrum shown in figure 6 was obtained by setting the magnetic
field to the peak of the 4.42 eV ODMR band and measuring the MCDA
spectrum over the range 3.11-6.00 eV with the microwave power on;
the microwave power was sufficiently attenuated so that no sample
heating was detectable. This measurement was then repeated with
the microwave power off; the difference between the two spectra
yielded the tagged MCDA spectrum.   No tagging data was available
above 6.00 eV due to the extremely high signal-to-noise ratio that
resulted after subtraction.  The preferred "digital lock-in"
tagging method described by Spaeth \cite{Ahlers}, which employs
low frequency modulation of the microwave power to induce changes
in the MCDA, was ineffective in this case, probably because of the
typically long spin-lattice relaxation times found in irradiated
glasses \cite{Bricis}. Since the tagging technique effectively
removes all diamagnetic signals from the measurement, it allowed
the bandshape of the Ge(1) center to be accurately measured
without concern of distortion from the GODC absorption. The tagged
MCDA spectrum for Ge(1) shows that the band extends from 3.5 eV to
approximately 5.3 eV.  Furthermore, the close similarity in
bandshape between the MCDA of figure 3 and the tagged spectrum
indicates that the Ge(1) center is responsible for all the
paramagnetic absorption in the range 2.01-5.3 eV; similar claims
may not be made for absorption bands above 5.3 eV using the
tagging technique due to high signal noise.  Attempts to tag the
MCDA spectra with the field set to the Ge(2) ESR line were
unsuccessful, as the signal noise was simply too great to produce
reliable data.  Thus, no further information about the nature of
the absorption above 5.3 eV could be obtained from these
experiments.

An obvious inconsistency with the standard defect model for
irradiated germanosilicate glasses emerges from the ODMR and
tagged MCDA results.  These techniques firmly establish that the
Ge(1) center is responsible for a broad absorption in the 4.4-4.7
eV range as had been previously suggested.  However, a discrepancy
is found regarding the optical properties of the Ge(2) center.
Recall that the Ge(2) defect is considered to be correlated with a
broad band centered in the 5.7-6.1 eV range, the optical
properties of which are determined by a gaussian fitting
procedure.  The gaussian fit to the optical absorption data shown
in figure 2b suggests that the 6.06 eV absorption ascribed to the
Ge(2) defect in the standard model has the largest peak oscillator
strength in the range measured, yet the MCDA spectrum clearly
shows considerably weaker paramagnetic absorption from 5.0-6.2 eV
than at 4.42 eV.  The lack of a strong paramagnetic signal in this
range indicates that the defect or defects responsible for the
majority of the optical absorption are likely diamagnetic in
nature; therefore, the Ge(2) defect does not correlate well with
the optical properties suggested by the standard model.  One
possible explanation for this is that the simple three-band
deconvolution of the induced optical absorption is inadequate, and
that one or more induced diamagnetic bands are responsible for the
majority of the optical absorption above 5 eV.  Further research
is currently underway to clarify these assignments, including
studies of UV-irradiated germanosilicate glasses; experiments with
irradiated hydrogen-loaded germanosilicates are also in progress.

\section{Conclusions}

Paramagnetic defects are created in 7.0 mol$\%$ Ge-doped SiO$_{2}$
glass via prolonged X-irradiation along with a number of optical
absorption bands in the range 2.01-6.21 eV.  ESR data suggests
Ge(1) and Ge(2) centers are formed in the greatest quantities,
while the GeE$'$ defect is experimentally undetectable.  MCDA
spectra of the irradiated glass measured at 4.2 K and 1.6K show a
large paramagnetic band at 4.42 eV, and much weaker paramagnetic
absorption above 5.0 eV.  Comparison of ODMR and ESR measurements
indicate that the 4.42 eV absorption is due to the Ge(1) center,
and tagged MCDA was used to positively identify all optical
transitions associated with this defect.  These results are in
accordance with the widely accepted model for Ge(1) in irradiated
germanosilicate glasses.  Lack of strong paramagnetic absorption
above 5.0 eV indicates that diamagnetic defects are responsible
for the bulk of the optical density in this region, contradicting
the widely accepted notion that the Ge(2) center is the dominant
absorbing center here. Comparison of ODMR and ESR data show that
the Ge(2) center may in fact absorb at 5.65 eV, which suggests its
peak oscillator strength lies in the deeper UV range of the
spectrum.  This result considered with the MCDA data indicate the
simple three-defect model in irradiated Ge-doped SiO$_{2}$ systems
must be augmented.

The authors gratefully acknowledge T. Erdogan and A. Heaney for
samples used in these experiments.  Work supported by NSF
DMR-9612267.

%
%

%

\begin{figure}
\caption{ESR spectra of X-irradiated Ge-doped silica glass
measured at 300 K.} \label{Figure 1}
\end{figure}

\begin{figure}
\caption{(a) Absorption spectra of unirradiated (dashed line) and
irradiated (solid line) Ge-doped silica glass.  (b) Radiation
induced change in optical absorption (solid line) with Gaussian
components (dashed lines).  Both sets of data were measured at 300
K.} \label{Figure 2}
\end{figure}

\begin{figure}
\caption{MCDA spectra of irradiated glass measured at 4.2 K
(dashed line) and 1.6 K (solid line) at 1.8 T.  Inset shows MCDA
spectrum of unirradiated glass measured at 4.2 K, 1.8 T.  X and Y
scales are identical on both plots.} \label{Figure 3}
\end{figure}

\begin{figure}
\caption{ODMR spectra measured at 4.42 eV (dashed line) and 5.65
eV (solid line) at 1.6 K.} \label{Figure 4}
\end{figure}

\begin{figure}
\caption{(a) ODMR spectrum measured at 4.42 eV, 1.6 K (solid line)
vs integrated ESR spectrum (dashed line) measured at 300 K.  (b)
ODMR spectrum measured at 5.65 eV, 1.6 K (solid line) vs
integrated ESR (dashed line).} \label{Figure 5}
\end{figure}

\begin{figure}
\caption{Tagged MCDA spectrum of Ge(1) center measured at 1.6 K.}
\label{Figure 6}
\end{figure}

%

\begin{references}
%
\bibitem{Atkins} R. M. Atkins and V. Mizrahi, Electron. Lett. {\bf28},
1743 (1992).

\bibitem{Hosono} H. Hosono, Y. Abe, D. L. Kinser, R. A. Weeks, K. Muta,
and H. Kawazoe, Phys. Rev. B {\bf46}, 11445 (1992).

\bibitem{Nishii} J. Nishii, K. Fukumi, H. Yamanaka, K. Kawamura, H. Hosono, and K.
Kawazoe, Phys. Rev. B {\bf52}, 1661 (1995).

\bibitem{Fujimaki} M. Fujimaki, K. Yagi, Y. Ohki, H. Nishikawa, and K. Awazu, Phys. Rev. B {\bf53},
9859 (1996).

\bibitem{Essid} M. Essid, J. Albert, J. L. Brebner, and K. Awazu, J.
Non-Cryst. Solids {\bf246}, 39 (1999).

\bibitem{Fujimakitwo} M. Fujimaki, T. Watanabe, T. Katoh, T. Kasahara, N. Miyazaki, Y. Ohki, and H. Nishikawa,
Phys. Rev. B {\bf57}, 3920 (1998).

\bibitem{Neustruev} V. B. Neustruev, J. Phys.:  Condens. Matter {\bf6}, 6901 (1994).

\bibitem{Shigemura} S. Shigemura, H. Y. Kawamoto, J. Nishii, and M. Takahashi, J. Appl. Phys. {\bf85}, 3413 (1999).

\bibitem{Tsai} T. E. Tsai, D. L. Griscom, and E. J. Friebele, Diff. Defect Data {\bf53-54}, 469 (1987).

\bibitem{Nishiitwo} J. Nishii, N. Kitamura, H. Yamanaka, H. Hosono, and H. Kawazoe,
Opt. Lett. {\bf20}, 1184 (1995).

\bibitem{Friebele} E. J. Friebele, D. L. Griscom, and G. H. Siegel Jr., J. Appl. Phys. {\bf45},
3424 (1974).

\bibitem{Anoikin} E. V. Anoikin, A. N. Guraynov, D. D. Gusovskii, V. M. Mashinskii, S. I. Miroshnichenko, V. B.
Neustruev, V. A. Tikhomirov, and Y. B. Zverev, Sov. Lightwave
Comm. {\bf1}, 123 (1991).

\bibitem{Hosonotwo} H. Hosono, H. Kawazoe, and J. Nishii, Phys. Rev. B {\bf53}, 11921 (1996).

\bibitem{Friebeletwo} E. J. Friebele and D. L. Griscom, in Defects In Glasses, eds. F. L. Galeener,
D. L. Griscom, and M. J. Weber (Materials Research Society,
Pittsburgh, 1985), p. 319.

\bibitem{Neustruevtwo} V. B. Neustruev, E. M. Dianov, V. M. Kim, V. M. Mashinsky, M. V. Romanov, A. N. Guryanov,
V. F. Khopin, and V. A. Tikhomirov, Fiber and Int. Optics {\bf8},
143 (1989).

\bibitem{Geschwind} S. Geschwind, Electron Paramagnetic Resonance (Plenum, New York, 1972), p. 386.

\bibitem{Spaeth} J. M. Spaeth and F. M. Lohse, J. Phys. Chem. Solids {\bf51},
861 (1990).

\bibitem{Stephens} P. J. Stephens, Adv. Chem. Phys. {\bf25}, 197 (1976).

\bibitem{Drake} A. F. Drake, J. Phys. E {\bf19}, 170 (1986).

\bibitem{Fujimakithree} M. Fujimaki, T. Katoh, T. Kasahara, N.
Miyazaki, and Y. Ohki, J. Phys.:  Condens. Matter {\bf11}, 2589
(1999).

\bibitem{Chiodini} N. Chiodini, F. Meinardi,, F. Morazzoni, A.
Paleari, and R. Scotti, Phys. Rev. B {\bf60}, 2429 (1999).

\bibitem{Guryanov} A. N. Guryanov, V. M. Kim, V. M. Mashinsky, V.
B. Neustruev, M. V. Romanov, V. A. Tikhomirov, and V. F. Khopin,
Proc. Gen. Phys. Inst. Moscow {\bf23}, 94 (1990).

\bibitem{Essidtwo} M. Essid, J. L. Brebner, J. Albert, and K.
Awazu, J. Appl. Phys. {\bf84}, 4193 (1998).

\bibitem{Anoikin} E. V. Anoikin, A. N. Guryanov, D. D. Gusovsky,
E. M Dianov, V. M. Mashinsky, S. I. Miroshnichenko, V. B.
Neustruev, V. A. Tikhomirov, and Y. B. Zverev, Nucl. Instrum.
Methods in Phys. {\bf B65}, 392 (1992).

\bibitem{Spaethtwo} J. M. Spaeth, Electron Magnetic Resonance of
the Solid State (The Canadian Society for Chemistry, Ottawa,
1987), p. 503.

\bibitem{Ahlers} F. J. Ahlers, F. Lohse, J. M. Spaeth, and L. F.
Mollenauer, Phys. Rev. B {\bf28}, 1249 (1983).

\bibitem{Bricis} D. Bricis, J. Ozols, U. Rogulis, J. Trokss, W.
Meise, and J. M. Spaeth, Solid State Communications {\bf81}, 745
(1992).

\end{references}
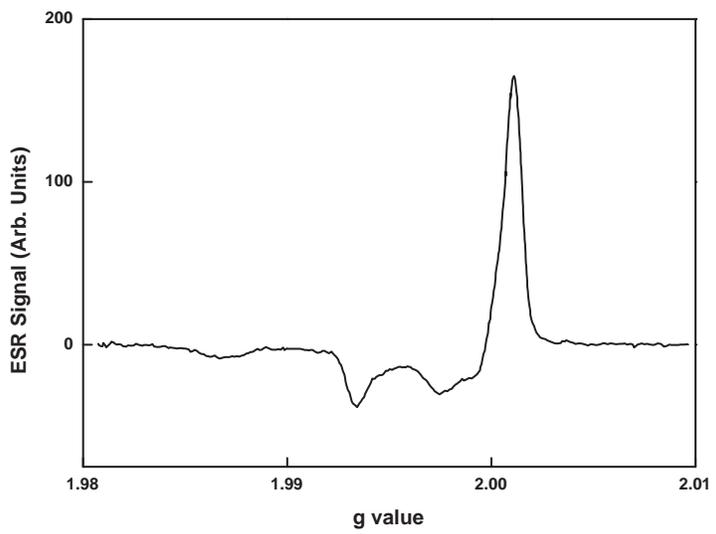
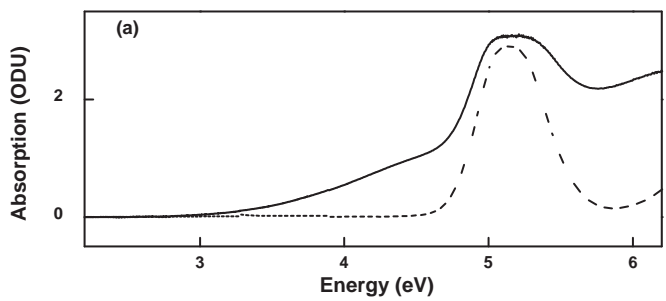
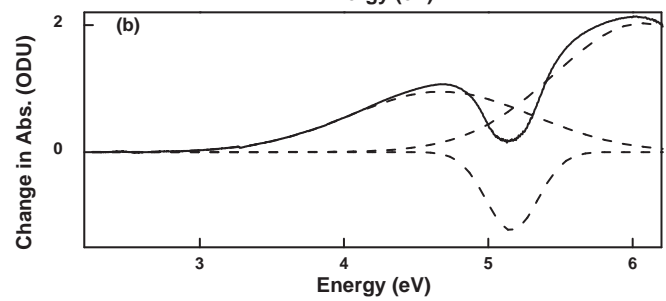
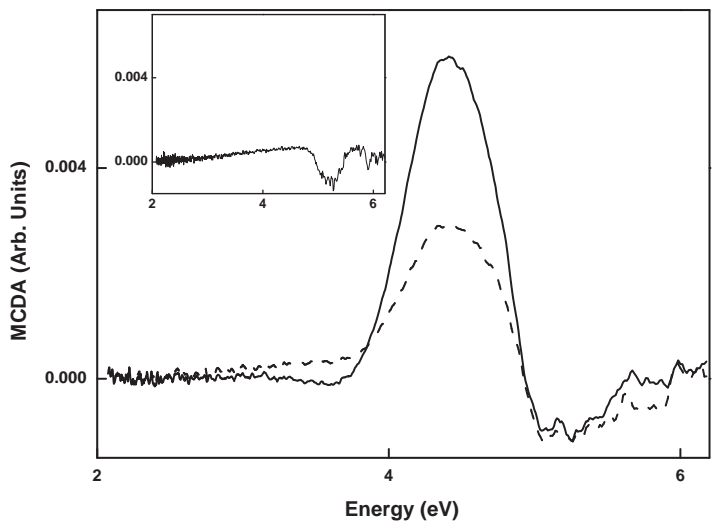
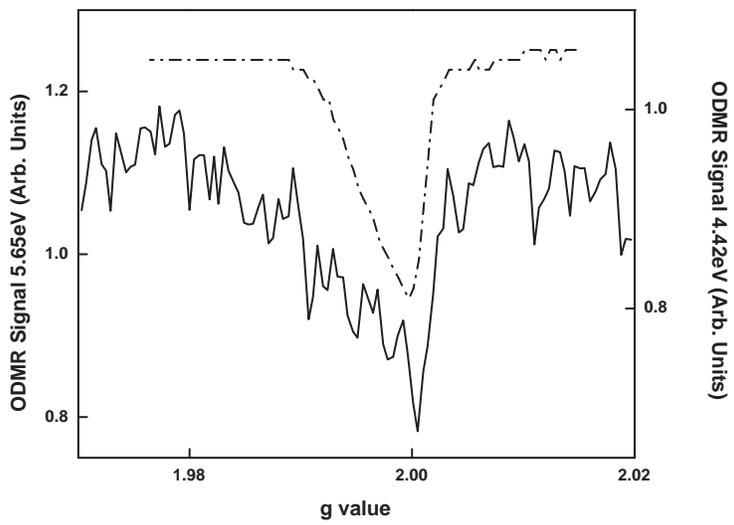
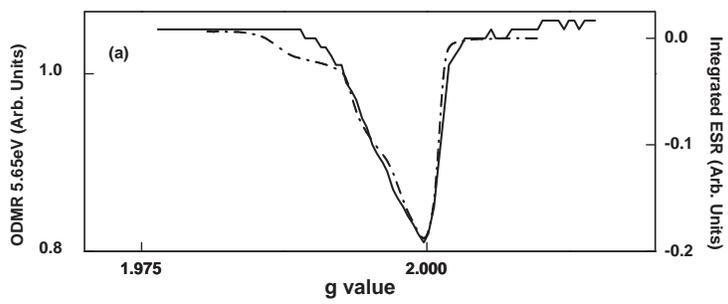
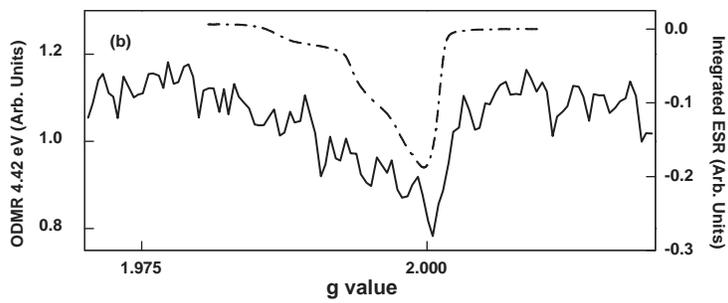
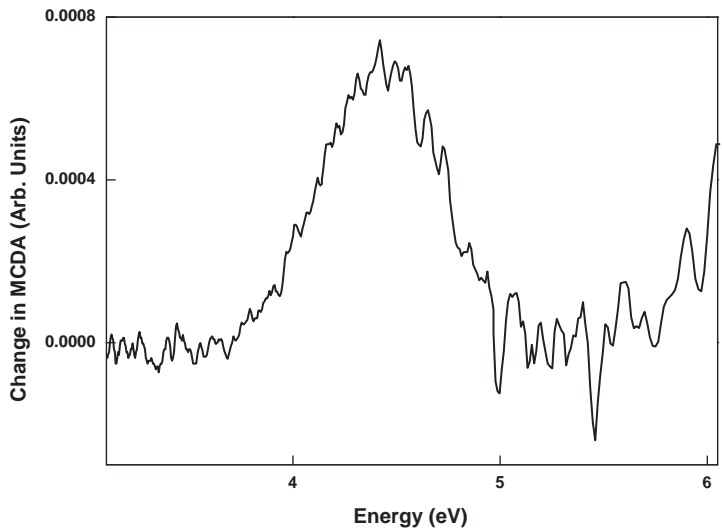
\end{document}